\documentclass[aps,prd,twocolumn,groupedaddress,amssymb,eqsecnum,showpacs,psfig]{revtex4}
\usepackage{graphicx}
\usepackage{bm}
\usepackage{dcolumn}
\usepackage{amsmath}

\begin{document}

\newcommand{\newc}{\newcommand}

\newc{\be}{\begin{equation}}
\newc{\ee}{\end{equation}}
\newc{\ba}{\begin{eqnarray}}
\newc{\ea}{\end{eqnarray}}
\newc{\bea}{\begin{eqnarray*}}
\newc{\eea}{\end{eqnarray*}}
\newc{\D}{\partial}
\newc{\ie}{{\it i.e.} }
\newc{\eg}{{\it e.g.} }
\newc{\etc}{{\it etc.} }
\newc{\etal}{{\it et al.}}
\newcommand{\nn}{\nonumber}

\newc{\ra}{\rightarrow}
\newc{\lra}{\leftrightarrow}
\newc{\no}{Nielsen-Olesen }
\newc{\lsim}{\buildrel{<}\over{\sim}}
\newc{\gsim}{\buildrel{>}\over{\sim}}
\title{Core Structure of Global Vortices in Brane World Models}
\author{L. Perivolaropoulos}
\email{leandros@cc.uoi.gr,
http://leandros.physics.uoi.gr} \affiliation{Department of
Physics, University of Ioannina, Greece}
\date{\today}

\begin{abstract}
We study analytically and numerically the core structure of global
vortices forming on topologically deformed brane-worlds with a
single toroidally compact extra dimension. It is shown that for an
extra dimension size larger than the scale of symmetry breaking
the magnitude of the complex scalar field at the vortex center can
dynamically remain non-zero. Singlevaluedness and regularity are
not violated. Instead, the winding escapes to the extra dimension
at the vortex center. As the extra dimension size decreases the
field magnitude at the core dynamically decreases also and in the
limit of zero extra dimension size we reobtain the familiar global
vortex solution. Extensions to other types of defects and gauged
symmetries are also discussed.

\end{abstract}

\maketitle

\section{Introduction}
The idea of extra dimensions has been an appealing concept in
theoretical physics since the 1920's when it was first considered
by Kaluza\cite{Kaluza:tu} and Klein\cite{Klein:tv}. They proposed
that our universe may be embedded in a five dimensional spacetime
where the fifth dimension is not detected because it is compact
and small. They achieved a unification of gravitational and
electromagnetic forces by inducing four dimensional gauge
interactions solely by the five dimensional geometry. Isometries
of the extra-dimensional compact manifold are then realized as
gauge symmetries of an effective four dimensional theory. In such
a framework the extra components of the metric play the role of
the gauge fields of the four dimensional theory.

Later, in the context of string theory (for a good review see
\cite{Antoniadis:1999yx}) it was shown that the existence of extra
dimensions is crucial not only for the unification of gravity with
other interactions but also for its quantization. String theory is
the only consistent quantum theory of gravity and its consistency
requires the existence of extra dimensions. Recently, the
possibility of having large extra dimensions has been considered
as a solution of the hierarchy
problem\cite{Antoniadis:1998ig,Arkani-Hamed:1998rs,
Arkani-Hamed:1998nn,Randall:1999ee,Randall:1999vf}. According to
this proposal, the fundamental Planck scale of the $4+D$
dimensions is close to the TeV scale not far from the electroweak
and strong interaction scales. The much larger value of the four
dimensional Plank mass $M_4$ would be due to the large size of the
extra dimensions according to the relation \be M_4^2 = b^D
M_{4+D}^{D+2} \ee where $b$ is the radius of the compact extra
dimensions. Consistency with phenomenology then requires that
whereas gravity can propagate in the $4+D$ dimensional bulk space,
the ordinary matter fields and gauge bosons be bound to live on a
3 dimensional brane which would constitute the usual spatial
dimensions.

The localization of the brane in the compact D-dimensional
manifold spontaneously breaks the corresponding Kaluza-Klein (KK)
isometries and defines a set of Goldstone boson fields which
parametrize the location of the brane in the compact
manifold\cite{Dvali:1999jd,Dvali:2000bz,Dobado:2000gr,Cembranos:2001my,Cembranos:2001rp}.
These zero modes can in general depend on the position on the
brane and their excitations have been called 'branons' in previous
studies\cite{Dobado:2000gr}.

Topologically non-trivial configurations of these geometrical
Goldstone fields can produce a new set of topological states:
vortices, monopoles\cite{Dvali:2000bz} and
skyrmions\cite{Cembranos:2001rp}. These defects form when the
isometry group $G$ of the D-dimensional compact manifold $M$
breaks down to a subgroup $H$ due to the localization of the
brane. If now the vacuum manifold $G/H$ has $n$-dimensional closed
surfaces that can not be contracted to a point in $G/H$
($\pi_n(G/H)\neq I$) then topologically non-trivial configurations
can form for $n=1$ (vortices), $n=2$ (monopoles) and $n=3$
(skyrmions). These form by mapping the surfaces $S^1$, $S^2$ or
$S^3$ from the physical space on the brane to the vacuum manifold.
They correspond to a constant change of the position of the brane
on the D-dimensional compact space (vacuum manifold) as we travel
a closed surface in 3+1 dimensions. For example for $n=1$, the
compact space is $S^1$ and a geometrical `cosmic string' forms
when travelling around such a string a four dimensional observer
makes a full circle along the extra dimension $S^1$. The condition
for such a geometrical string to form is \be \theta_2 (\theta_1) =
\theta_1 \ee where the Goldstone field $\theta_2$ parametrizes the
position of the brane along the extra dimension $S^1$ while
$\theta_1$ is the usual azimuthal angle on the brane. On the other
hand no string would form if $\theta_2 = const$ and the Goldstone
field is independent of position on the brane.

It has been shown\cite{Dvali:2000bz} that in the core of such
defects, the broken isometries of the compact space get restored
and the four dimensional observer is free to move anywhere along
the D-dimensional compact space. Another interesting possibility
which has not been investigated so far is the modification of the
size of the extra dimensions at the geometrical defect core due to
energetic reasons (the stabilization of extra dimensions size is
dynamical). This could lead to spatially variable size of extra
dimensions with interesting cosmological
consequences\cite{Morris:2002sd,Guendelman:2003jg}.

These topological defects are geometrical since they form due to
the spontaneous breaking of isometries of extra dimensions and are
distinct from the usual defects formed by the spontaneous breaking
of non-geometrical scalar field symmetries. However,
non-geometrical defects do admit several new generalized
properties in the context of brane-world models. For example the
3+1 dimensions of our brane-world can be identified with the core
of topological defects residing in a higher dimensional space-time
(a domain wall in 5d\cite{Rubakov:bb,KorthalsAltes:2001et}, string
in
6d\cite{Cohen:1999ia,Gregory:1999gv,Olasagasti:2000gx,Gherghetta:2000qi,Gregory:2002tp,Cline:2003ak},
monopole in 7d\cite{Gherghetta:2000jf,Benson:2001ac} or other
defects\cite{Demir:2000gj,Cho:2002ui}). In this framework, the
localization of matter fields on the defect core can be associated
with the fermionic and scalar zero modes localized in defect
cores\cite{Giovannini:2002mk}. The scalar fields forming these
defects are assumed to propagate in the extra dimensions (the
bulk) and depend only on the extra coordinates. For example, the
scalar field of a global vortex  in 6d would be \be \phi(\rho_2,
\theta_2)=f(\rho_2) \; e^{i m \theta_2} \ee where $\rho_2$ and
$\theta_2$ are the bulk radius and bulk angle and $m$ is the
integer winding number (in what follows we set $m=1$ for
simplicity).

In addition to this brane-world generalization of non-geometrical
defects we can consider another interesting generalization using
just scalar fields that are localized on the brane
$(t,\rho,\theta_1,z)$. Consider for example the brane localized
complex scalar field ansatz: \be \label{anz} \phi (\rho, \theta_1,
\theta_2)= f(\rho) \; e^{i\alpha \theta_1 + i(1-\alpha)\theta_2}
\ee where $\phi$ is localized on the $3+1$ brane of a 5d bulk with
a toroidally compact extra dimension of size $b$ parametrized by
$\theta_2$ and $\alpha(\rho)$ is a real function of $\rho$ with
\be \label{bc} \alpha(\rho \rightarrow \infty)=1 \ee for unit
winding vortex. If the location of the brane in the bulk is
independent of the position on the brane, the branons are not
excited (locally or topologically) and we may set $\theta_2 =
const$. In this case, regularity of $\phi$ at the origin $\rho=0$
implies that \be \label{cc}  \alpha(\rho) =1 \;\;\; f(0)=0 \ee and
we obtain the well known global vortex.

A much more interesting case appears when we have topological
excitations of branons. The simplest such case occurs when
$\theta_2 = \theta_1$. In this case, the boundary condition
(\ref{bc}) does not imply the usual conditions (\ref{cc}).
Instead, it is dynamics that will determine the core structure of
the global vortex. This will be demonstrated in the next section.
It will be seen that the global vortex generalizes in this case
with a new dimensionless parameter ${\bar b}\equiv \eta b$ which
is the ratio of the size of the compact extra dimension over the
scale of symmetry breaking of the vortex $\eta^{-1}$. For $\eta \;
b \rightarrow 0$ we dynamically reobtain the usual core structure
of the global vortex ($f(0)=0$). For ${\bar b} \gg 1$ we have a
new structure with $f(0) \simeq \eta$ without violation of
regularity since it is energetically favorable in this case for
the winding to escape in the extra dimension with $\alpha(0) = 0$.
The gauged case and the extension to other types of defects is
discussed in section III. Finally in section IV we conclude and
discuss the possible extensions and implications of these results.

\section{The Global Brane Vortex}
Consider a spacetime with topology ${\cal M}_4 \times S^1$ where
${\cal M}_4$ is our four dimensional Minkowski space and $S^1$ is
a torroidally compact extra dimension. In this spacetime we
consider a Lagrangian of a complex scalar field $\phi$ with a
broken global $U(1)$ symmetry \be {\cal L} = {1\over 2} \partial_M
\phi^*
\partial^M \phi - {\lambda \over 4} (\phi^* \phi - \eta^2)^2 \ee
where $M=0,...,4$ and the extra coordinate $x^4$ is compact with
size $b$ and parametrized by an angle $\theta_2$. The
corresponding energy density of a static scalar field
configuration is of the form \be {\cal E} = {1\over 2} \vert
{{\partial \phi}\over {\partial \rho}}\vert ^2 +{1\over 2} {1\over
\rho^2} \vert {{\partial \phi}\over {\partial \theta_1}}\vert ^2 +
{1\over 2} \vert {{\partial \phi}\over {\partial z}}\vert ^2 +
{1\over 2} {1\over b^2} \vert {{\partial \phi}\over {\partial
\theta_2}}\vert ^2 + {\lambda \over 4} (\phi^* \phi - \eta^2)^2
\ee where $(\rho, \theta_1,z)$ are the usual cylindrical
coordinates of $M_4$.

We now use the generalized global vortex ansatz (\ref{anz}) with
boundary conditions at infinity \be \label{bcinf}  \alpha(\rho)
\rightarrow 1 \;\;\; f(\rho)\rightarrow \eta \ee The energy
density for this ansatz takes the form \be \label{edens} 2 {\cal
E}=f'^2 + \alpha'^2 (\theta_1 - \theta_2)^2 + {{\alpha^2 f^2}\over
\rho^2} + {{(1-\alpha)^2 f^2}\over b^2}+{\lambda \over 4} (f^2 -
\eta^2)^2 \ee We now distinguish between two topologically
inequivalent brane configurations
\begin{enumerate} \item The vacuum sector with $\theta_2 =
constant$ where all points of the brane are located at the same
extra coordinate. In this case continuity and regularity of $\phi$
at the origin $\rho =0$ implies that $f(0)=0$ and $\alpha
(\rho)=1$. We thus obtain the usual global vortex. \item The
topological sector with $\theta_2 = \theta_1$ where as we span a
circle on the brane we at the same time span a circle in the extra
dimension. Thus the brane is deformed along the extra dimension in
a topologically non-trivial way. In this case the core properties
of the global vortex can be modified in two important
ways.\begin{itemize} \item The function $\alpha(\rho)$ is not
required to be integer and constant. Regularity and
singlevaluedness of $\phi$ are not violated for any value of
$\alpha(\rho)$. Thus $\alpha(\rho)$ is determined by dynamics
(energy minimization). \item If dynamical arguments lead to
$\alpha(0)=0$ then $\phi$ can be regular at the origin with
$f(0)=c\neq 0$. This is a novel type of behavior for the global
vortex core. \end{itemize}
\end{enumerate} In what follows we focus on the case of the
topological sector ($\theta_2 = \theta_1$) and investigate the
properties of the global vortex by deriving the field functions
$\alpha(\rho)$ and $f(\rho)$ with the boundary conditions
(\ref{bcinf}). The analytical derivation of $\alpha(\rho)$ is
simple since for $\theta_2 = \theta_1$ the energy density
(\ref{edens}) is independent of $\alpha'$. Thus we simply demand
\be {{\delta {\cal E}}\over {\delta \alpha}}=0 \ee which leads to
\be \label{aform} \alpha(\rho) = {\rho^2 \over{b^2 + \rho^2}} \ee
Thus $\alpha(\rho)$ has the expected asymptotic behavior at $\rho
\rightarrow \infty$ ($\alpha \rightarrow 1$) and at $\rho
\rightarrow 0$ ($\alpha \rightarrow 0$). This dynamical behavior
of $\alpha(\rho)$ implies that the winding at the origin transfers
from the brane coordinate $\theta_1$ to the extra dimension
$\theta_2$. Since now $\phi$ is independent of $\theta_1$ at the
origin $\rho =0$, it does not have to be zero to maintain its
singlevaluedness and regularity. Thus, the boundary condition for
$f(\rho)$ at the origin is $f(0)=c$ (and $f'(0)=0$ by symmetry
considerations) where the values of $c$ and $f(\rho)$ are to be
determined dynamically in terms of the single dimensionless
parameter of the model $ {\bar b}= \eta b$. In order to find
$f(\rho)$ we rewrite the energy density (\ref{edens}) in
dimensionless form using $\theta_1 =
\theta_2 $. After setting \ba {f\over \eta}&\rightarrow & f \\
\lambda \eta^2 \rho^2 &\rightarrow & \rho^2 \\ \theta_1 &=&
\theta_2 \\ \lambda \eta^2 b^2 &\rightarrow & b^2 \ea equation
(\ref{edens}) becomes \be \label{redens} {{2 {\cal E}}\over
{\lambda \eta^2}} \equiv {\bar{\cal E}}=f'^2  + {{\alpha^2
f^2}\over \rho^2} + {{(1-\alpha)^2 f^2}\over b^2}+{1 \over 2} (f^2
- 1)^2 \ee The rescaled field equation for $f(\rho)$ is \be
\label{feq} f'' + {f' \over \rho} - {{f \alpha^2}\over \rho^2} -
{{f(1-\alpha)^2}\over b^2} -(f^2-1)f=0 \ee For $b\rightarrow 0$
equation (\ref{aform}) gives $\alpha =1$ and we obtain the usual
global vortex field equation as we should. For $b\neq 0$ we may
use equation (\ref{aform}) to eliminate $\alpha(\rho)$ from
(\ref{redens}) and (\ref{feq}) leading to \be \label{refin}
{\bar{\cal E}} = f'^2  + {{f^2}\over {\rho^2 + b^2}} +{1 \over 2}
(f^2 - 1)^2 \ee for the rescaled energy density and \be
\label{feqfin} f'' + {f' \over \rho} - {{f}\over
{\rho^2+b^2}}-(f^2-1)f=0 \ee for the rescaled field equation. For
small $\rho$ ($\rho<1$ and $\rho<b$) equation (\ref{feqfin})
implies $f(\rho)=c$ where $c$ is a constant. For $b<1$ and in the
range $b<\rho < 1$, $f(\rho)$ is approximated by \be \label{fapx}
f(\rho)=c\;{\rho \over b} \ee By extrapolating equation
(\ref{fapx}) up to $\rho = {b\over c}$ and setting $f=1$ beyond
that, we obtain a crude analytical approximation for the field
function $f(\rho)$. This analytic approximation can be used to
minimize the energy with respect to $c$ for various values of $b$.
Thus, finding the value of $c$ that minimizes the energy for a
given $b$ we have a crude analytic approximation of $c(b)$ to
compare with the corresponding numerical result obtained by
solving the diffrential equation (\ref{feqfin}) with boundary
conditions $f(\infty)=1$ and $f'(0)=0$. This comparison is shown
in Fig. 1.
\begin{figure}
\centering
\includegraphics[height=8cm,angle=-90]{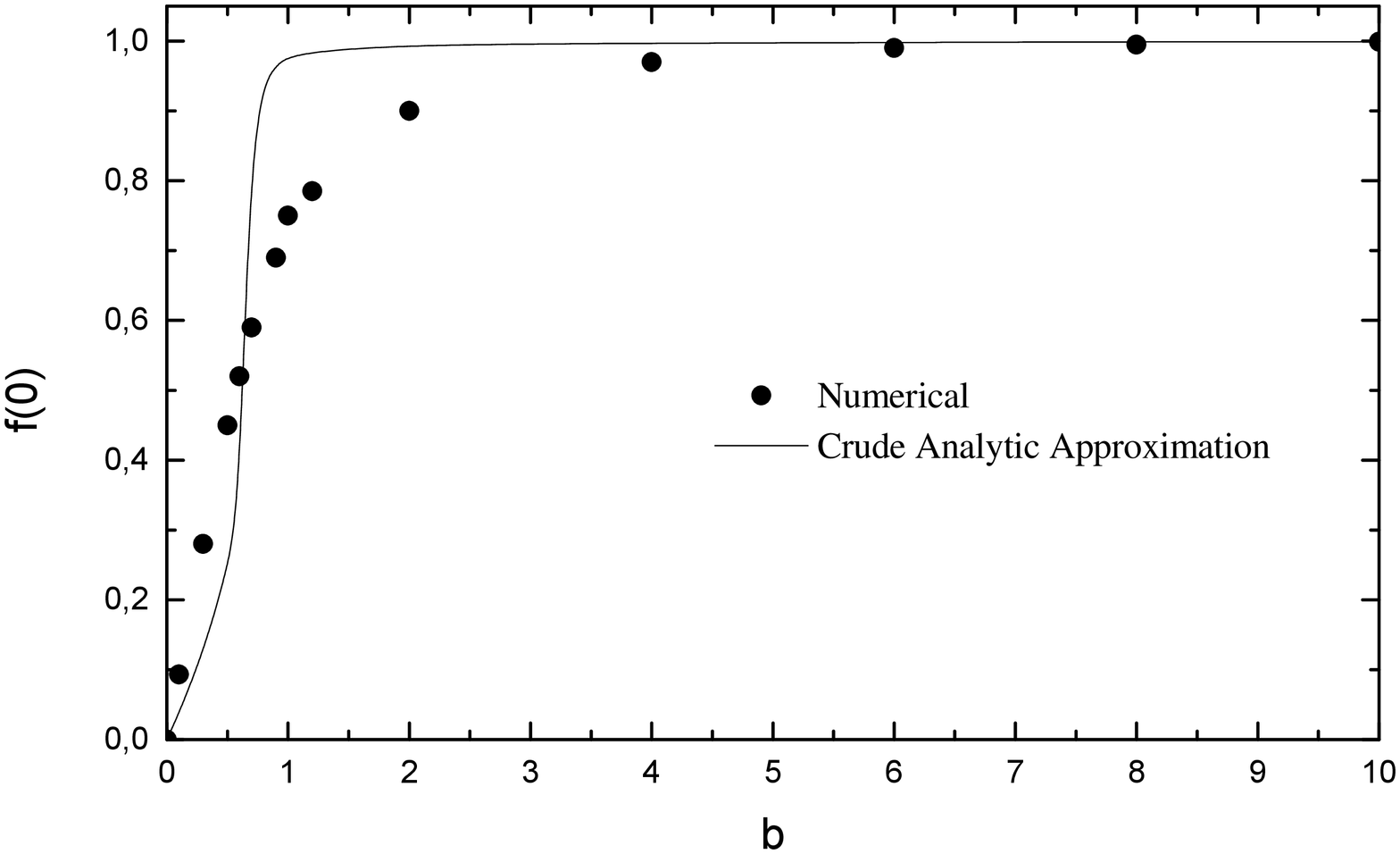}
\caption{The field of the global vortex is non-zero at the origin
for a non-zero size $b$ of the extra dimension.} \label{fig1}
\end{figure}
The corresponding forms of the field function $f(\rho)$ obtained
numerically by solving equation (\ref{feqfin}) for various values
of $b$ are shown in Fig. 2.
\begin{figure}
\centering
\includegraphics[height=8cm,angle=-90]{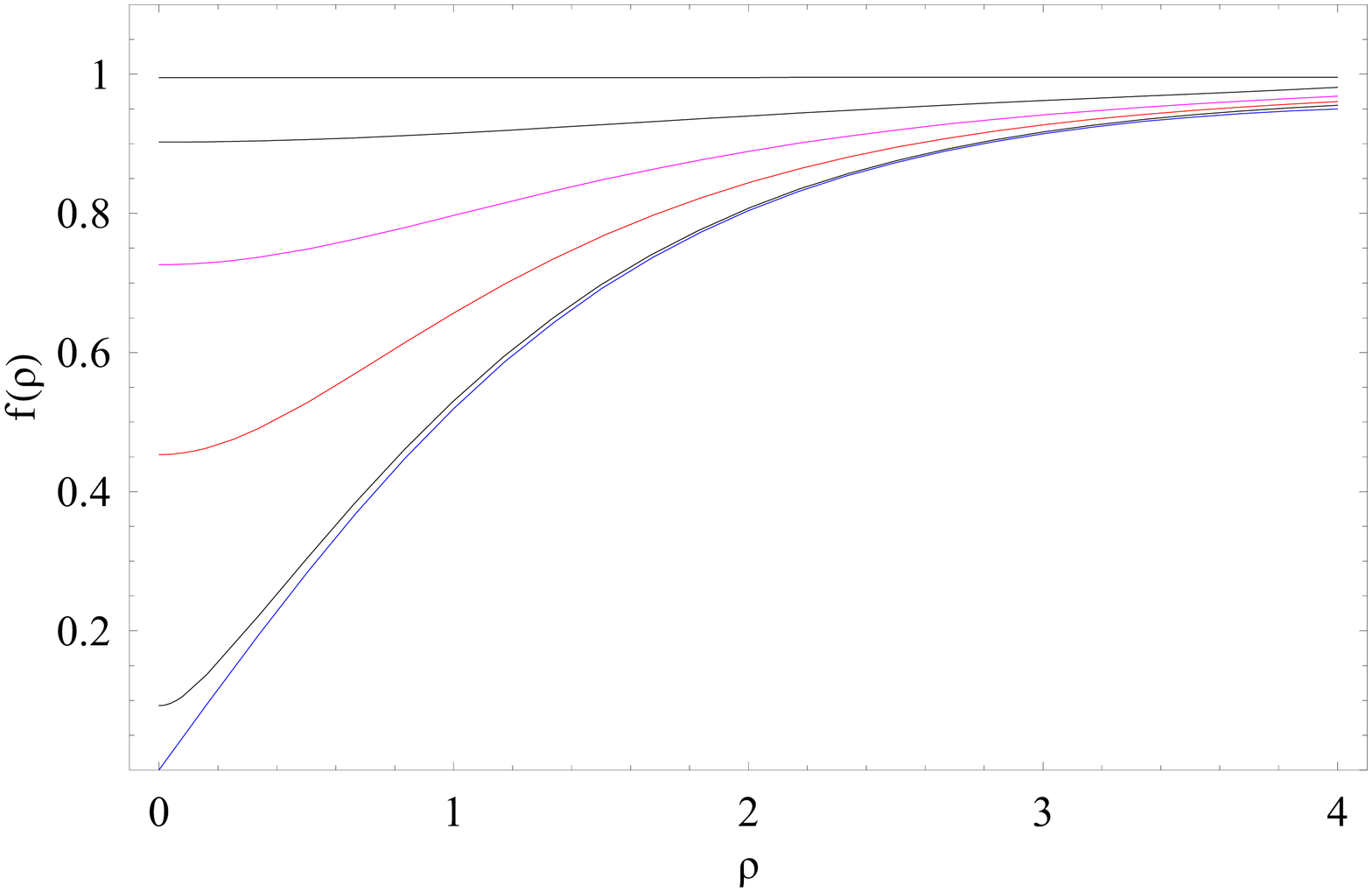}
\caption{The profile $f(\rho)$ of the global vortex field for
$b=0,\; 0.1,\; 0.5,\; 1.0,\; 2.0,\; 10.0$. As expected $f(0)$
increases with $b$ and $f(\rho)$ remains in the vacuum for large
$b$.} \label{fig2}
\end{figure}
As expected, for a size of the extra dimension much less than the
scale of symmetry breaking ($b\ll 1$) $f(0)$ goes continously to
$0$ and the usual global vortex solution is reached. At the other
extreme where the extra dimension size is much larger than the
scale of symmetry breaking  ($b\gg 1 $) the energy density
approaches $0$ everywhere. In this limit, $\vert \phi \vert$
remains at its vacuum everywhere while the winding is smoothly
transferred to the extra dimension with practically no cost in
gradient energy. This transition is shown in Fig. 3 where the
numerically obtained energy density is shown to smoothly approach
$0$ on scales $\rho < b$ for large enough compactification radius.
\begin{figure}
\centering
\includegraphics[height=8cm,angle=-90]{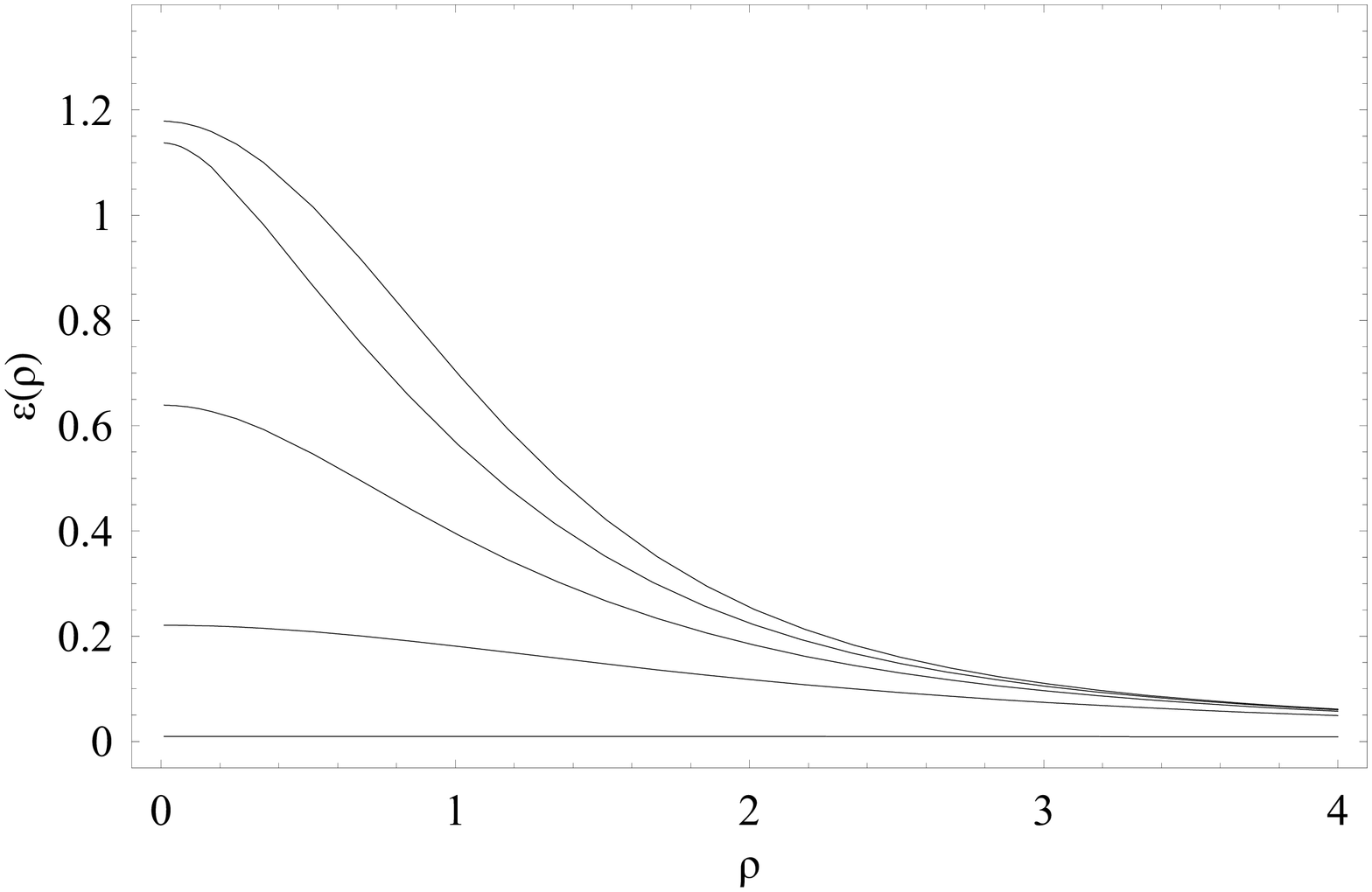}
\caption{The energy density of the vortex decreases for increasing
size $b$ of the extra dimension. The densities corresponding to
$b=0,\; 0.5,\; 1.0,\; 2.0,\; 10.0$ are shown. }\label{fig3}
\end{figure}
However, the total energy of the global vortex is \be \label{etot}
{\cal E} \sim \int_0^{R^2} {{d\rho^2}\over {b^2 +\rho^2}} \sim
ln({R\over b}) \ee and diverges logarithmically for any finite
$b$. It is worth noting that the escape of the winding and of the
scalar field to the extra dimension at the string core is
consistent with the results of Ref. \cite{Dvali:2000bz} where it
was shown that the center of a topologically deformed brane is a
window to the extra dimensions for all fields and particles.

\section{Generalized Nielsen-Olesen Vortex}
In this section we attempt a generalization of the
Nielsen-Olesen\cite{Nielsen:cs} (NO) vortex along the lines of the
previous section where the global vortex was generalized by an
embedding to topologically non-trivial brane-worlds. The energy
density of the static Abelian Higgs model embedded in a
brane-world with a toroidally compact extra dimension is \be
\label{noedens1} {\cal E}=(D_i \phi)^* (D_i \phi) + {\lambda \over
2} (\phi^* \phi - \eta^2)^2 + {1\over 2} F_{ij} F_{ij} \ee where
$D_i = \partial_i - i e A_i$ and $F_{ij}=\partial_i A_j -
\partial_j A_i$ with $i,j=1,2,3$. The spatial coordinate $x_3$ is
taken to be toroidally compact ($x_3 = b \theta_2$). In analogy
with the usual NO vortex in two spatial dimensions and with the
results of the previous section we use the following
generalization of the NO ansatz \ba \phi &=& f(\rho) e^{i(\alpha
\theta_1 + (1-\alpha)\theta_2)} \\
A_1 &=& - a_1(\rho) \sin\theta_1 \\
A_2 &=& a_1(\rho) \cos \theta_1 \\
A_3 &=& a_2 (\rho) \ea Using this ansatz and equation
(\ref{noedens1}) we may express the energy density in terms of
$f(\rho)$, $a_1(\rho)$ and $a_2(\rho)$. The result (for $\theta_2
= \theta_1$) in dimensionless form is \ba  {{2 {\cal E}}\over
{\eta^4 e^2}} \equiv {\bar{\cal E}} = f'^2 + {f^2 \over \rho^2}
(\alpha - a_1)^2 + && \nn \\ \label{noredens1}  +{f^2 \over b^2}
((1-\alpha) - a_1)^2 + {\beta \over 2} (f^2 -1)^2 + {a_1'^2 \over
\rho^2} + {a_2'^2 \over b^2}&& \ea where we have set
\ba {f\over \eta} &\rightarrow &  f \\
e a_1&\rightarrow & a_1 \\
e a_2&\rightarrow & a_2  \\
e \eta \rho &\rightarrow & \rho \\
e \eta b &\rightarrow & b \\
{\lambda \over e^2} &\rightarrow & \beta \ea As in the global case
the functional form of $\alpha (\rho)$ is obtained analytically by
energy minimization \be \label{noar} {{\delta {\cal E}}\over
{\delta \alpha}} = 0 \Rightarrow \alpha = {{\rho^2 + b^2 a_1 -
\rho^2 a_2}\over {b^2 + \rho^2}} \ee which has the correct global
limit. Using equation (\ref{noar}) for $\alpha(\rho)$ in
(\ref{noredens1}) we may write the rescaled energy density as \be
\label{noredens2} {\bar {\cal E}}= f'^2 + {{f^2 (1-a_1
-a_2)^2}\over {\rho^2 + b^2}}
 + {\beta \over 2} (f^2 -1)^2 + {a_1'^2 \over \rho^2}+ {a_2'^2 \over b^2}
\ee By extremizing ${\bar {\cal E}}$ with respect to $f$, $a_1$
and $a_2$ we find the corresponding field equations \ba
 f'' + {f' \over \rho} - {{(1-a_1-a_2)^2}\over {b^2
+ \rho^2}} f - \beta (f^2 -1) f = 0 && \nn \\ \label{norsyst}
a_1'' - {a_1' \over \rho} +{{(1-a_1-a_2) \rho^2}\over {b^2 +
\rho^2}} f^2  = 0 && \\
a_2'' + {a_2' \over \rho} +{{(1-a_1-a_2) b^2}\over {b^2 + \rho^2}}
f^2 = 0 && \nn \ea The boundary conditions at infinity are \be
f\rightarrow 1, \; \; a_1'\rightarrow 0 \; \; a_2' \rightarrow 0
\ee and at the origin \be f' \rightarrow 0, \; \; a_1 \rightarrow
0 \; \; a_2' \rightarrow 0 \ee The next step would be to solve the
system (\ref{norsyst})  with the above boundary conditions to
obtain the field functions $f(\rho)$, $a_1 (\rho)$ and $a_2
(\rho)$. However, the existence of a trivial zero energy solution
satisfying the boundary conditions saves the effort. The solution
is \be f=1 \;\; a_1 = 0 \;\; a_2 = 1 \ee ie there is a
non-singular gauge transformation that transforms the vortex to a
$0$ energy configuration. Notice that we are free to take $a_2
\neq 0$ at $\rho =0$ without violating singlevaluedness of the
gauge field ${\vec A}$ since ${\vec A}$ does not wind along the
extra dimension and also $b\neq 0$. We conclude that the gauged
analog of the generalized global vortex discussed in the previous
section is trivial.

The brane-world defect generalizations discussed in this and in
the previous section can be extended to the cases of monopoles and
skyrmions. For example the case of global
monopoles\cite{Barriola:hx} would require two extra dimensions
with spherical compactification parametrized by the angles
($\theta_2, \varphi_2$). The corresponding generalized global
monopole ansatz is \be
\phi(\rho,\theta_1,\varphi_1,\theta_2,\varphi_2)=
f(\rho)\left(\begin{array}{c} \sin\Theta \cos
\Psi \\
\sin\Theta \sin\Psi\\
\cos\Theta
\end{array} \right) \ee
where \ba \Theta & \equiv &  \alpha_\theta \theta_1 +
(1-\alpha_\theta)\theta_2 \\
\Psi &\equiv & \alpha_\varphi \varphi_1 +
(1-\alpha_\varphi)\varphi_2 \ea $\theta_1$, $\varphi_1$ are the
usual spherical coordinates on the brane and $\alpha_\theta
(\rho)$, $\alpha_\varphi (\rho)$ generalize the function $\alpha
(\rho)$ of equation (\ref{anz}). The detailed study and
classification of all generalized brane-world defects is an
interesting open issue.

\section{Conclusion - Outlook}
We have studied field configurations of vortices in braneworlds
which are topologically deformed along a bulk with a toroidally
compact extra dimension. We have found that the field
configuration of generalized global vortices depends on a single
dimensionless free parameter ${\bar b}=\eta b$: the ratio of the
size of the extra dimension $b$ over the scale of symmetry
breaking $\eta^{-1}$ that gives rise to the vortex. For $\eta b
\rightarrow 0$ the effects of the extra dimension can be ignored
and the field configuration is that of the well known global
vortex. For $\eta b \simeq O(1)$ the core of the vortex acquires
novel properties: the complex scalar field remains non-zero and
singlevalued at the origin while its winding escapes to the extra
dimension. In addition, the energy density of the vortex gets
reduced at the origin since the potential (and gradient) energies
get smaller. For $\eta b \gg 1$ the energy density is negligible
on scales $\rho < b$ since the winding resides exclusively on the
large extra dimension on these scales, contributing negligible
gradient energy. Thus it is possible to have topologically
non-trivial configurations with negligible energy density at the
core. The total energy however was shown to remain divergent
logarithmicaly like $ln{R\over b}$. It should be noted that for
large extra dimensions (${\cal O}(\geq 1nm)$) and vortices
produced at symmetry breaking energy scales larger than the
electroweak scale we anticipate $\eta b \gg 1$.

For the gauged case we showed that the extra component of the
gauge field corresponding to the extra dimension is able to
eliminate the energy density of the generalized NO vortex even
though its topological properties remain non-trivial\footnote{A
similar effect happens when one tries to gauge global textures}.
If however only the coordinates on the brane are gauged then the
energy density of the vortex can not be eliminated and its core
properties will be similar to those of the brane-world global
vortex. The extension of this study to other types of defects in
topologically non-trivial brane-worlds is an issue worth of
further investigation.

{\bf Acknowledgements:} I thank I. Olasagasti for interesting
conversations. This work was supported by the European Research
and Training Network HPRN-CT-2000-00152.

\end{document}